\documentclass[preprint,dvips,prd,showpacs]{revtex4}
\usepackage{amsfonts}
\usepackage{amsmath,amssymb,epsf,epsfig,amsthm,bm,graphicx,subfigure}

\input{tcilatex}

\begin{document}

\title{\begin{flushright}
{ \small CECS-PHY-08/10 }
\end{flushright}\vskip1.0cm General Relativity with small cosmological
constant\\
from spontaneous compactification of Lovelock theory in vacuum}
\author{Fabrizio Canfora$^{1}$, Alex Giacomini$^{2,1}$, Ricardo Troncoso$%
^{1,3}$ and Steven Willison$^{1}$}
\email{canfora, giacomini, troncoso, steve@cecs.cl}
\affiliation{$^{1}$Centro de Estudios Cient\'{\i}ficos, Casilla 1469, Valdivia, Chile.}
\affiliation{$^{2}$Instituto de F\'{\i}sica, Facultad de Ciencias, Universidad Austral de
Chile, Valdivia, Chile}
\affiliation{$^{3}$Centro de Ingenier\'{\i}a de la Innovaci\'{o}n del CECS (CIN),
Valdivia, Chile.}
\pacs{04.50.-h, 04.50.Cd, 04.50.Kd}

\begin{abstract}
It is shown that Einstein gravity in four dimensions with small cosmological
constant and small extra dimensions can be obtained by spontaneous
compactification of Lovelock gravity in vacuum. Assuming that the extra
dimensions are compact spaces of constant curvature, General Relativity is
recovered within certain class of Lovelock theories possessing necessarily
cubic or higher order terms in curvature. This bounds the higher dimension
to be at least seven. Remarkably, the effective gauge coupling and Newton
constant in four dimensions are not proportional to the gravitational
constant in higher dimensions, but shifted with respect to their standard
values. This effect opens up new scenarios where a maximally symmetric
solution in higher dimensions could decay into the compactified spacetime
either by tunneling or through a gravitational analogue of ghost
condensation. Indeed, this is what occurs requiring both the extra
dimensions and the four-dimensional cosmological constant to be small.
\end{abstract}

\maketitle

\section{Introduction}

The original dream of unifying the fundamental interactions through the
dimensional reduction of pure General Relativity (\textbf{GR}) with
cosmological constant in vacuum is appealing not least because of its
simplicity and economy of ingredients. However, requiring both the inclusion
of nonabelian fields and a small four-dimensional cosmological constant, is
inconsistent with the fact that the compact \textquotedblleft
internal\textquotedblright\ manifold has to be of sufficiently small size
(see e.g., \cite{OW}). Thus, if one keeps GR as the higher-dimensional
theory to begin with, one cannot avoid introducing new fundamental matter
fields in higher dimensions. Albeit that this fact somehow perverts the
original dogma of obtaining bosonic fields from pure geometry in higher
dimensions, matter fields of a higher-dimensional origin could be welcome in
this vein insofar as the arbitrariness in their choice were removed by
requiring some basic principle for the fundamental theory to hold; as for
instance, local supersymmetry. Most of what we have learned about
dimensional reduction has been developed along this line, mainly within the
context of supergravity in diverse dimensions and string theory (For a
review see, e.g., \cite{DK07} and references therein).

We propose to explore whether one could return to the original dream of
having a realistic model for bosonic fields arising from pure geometry in
higher dimensions. This scenario should then accommodate consistently
nonabelian gauge fields and small cosmological constant with a sufficiently
small compact internal manifold.

Note that in dimensions higher than four, GR is not the only option for the
gravity theory to begin with. Indeed, the same basic requirements that yield
the Einstein-Hilbert action with cosmological constant in four dimensions,
i.e., general covariance and second order field equations for the metric,
give rise to the Lovelock action \cite{Lovelock}. This action contains
higher powers in the curvature in a precise combination and its Kaluza-Klein
reduction will always give field equations for the matter fields of second
order. The simplest modification of GR in higher dimensions corresponds to
the addition of quadratic terms in the so-called Gauss-Bonnet combination.
Thus, it is natural wondering whether the Einstein-Gauss-Bonnet (\textbf{EGB}%
) theory admits suitable spontaneous compactifications to four dimensions.
In the case of spontaneous compactifications of EGB theory in vacuum, the
Einstein equations are recovered, but with an additional scalar condition
which overconstrains the four-dimensional gravitational field, so that even
the Schwarzschild solution becomes excluded \cite{GOT}.

Nonetheless, here it is shown that GR with small cosmological constant,
without any additional constraint on the four-dimensional geometry, can be
recovered from spontaneous compactifications with small extra dimensions,
provided cubic or higher order terms in the Lovelock series are considered,
with a single relation amongst the coupling constants. Remarkably, the
effective four-dimensional Newton constant is not proportional to the
gravitational constant in higher dimensions, but instead shifted with
respect to the standard value. This effect opens up new interesting
scenarios which are analyzed below. The mechanism bounds the higher
dimension to be at least seven, since this is the smallest dimension that
admits a nontrivial cubic term in the action. It is instructive to perform
the complete analysis in this case.

\section{Spontaneous compactifications.}

In seven dimensions, the Lovelock action acquires a simple expression when
it is written in terms of differential forms, which reads\vspace*{-0.12in}

\begin{align}
I_{7}& =\int_{M_{7}}\!\!\!\!\!\epsilon _{ABCDEFG}\left(
\!c_{3}R^{AB}R^{CD}R^{EF}\!\!+\!\frac{c_{2}}{3}R^{AB}R^{CD}e^{E}e^{F}\right.
\notag \\
& \left. \!\!+\frac{c_{1}}{5}R^{AB}e^{C}e^{D}e^{E}e^{F}\!\!+\!\frac{c_{0}}{7}%
e^{A}e^{B}e^{C}e^{D}e^{E}e^{F}\right) \!e^{G}\,,  \label{Action7}
\end{align}%
where $R^{AB}$ and $e^{A}$ stand for the curvature two-form and the
vielbein, respectively, and wedge product between forms is understood. Let
us look for spontaneous compactifications of the form $M_{7}=M_{4}\times
K_{3}$, where $M_{4}$ is a four-dimensional Lorentzian manifold, and $K_{3}$
is a compact Euclidean manifold, which for simplicity is assumed to be of
constant curvature $R^{ab}=\Lambda _{3}e^{a}e^{b}$. Indices are split into
greek on $M_{4}$ and lowercase Latin on $K_{3}$. Then, it is simple to show
that in the vanishing torsion sector, the field equations along $M_{4}$
reduce to the Einstein equations with cosmological constant in vacuum 
\begin{equation}
\frac{1}{8\pi \tilde{G}_{4}}\left( G_{\mu \nu }+\Lambda _{4}g_{\mu \nu
}\right) =0\ ,  \label{efen11}
\end{equation}%
where the effect of the quadratic and cubic terms just amounts to a
redefinition of the Newton and cosmological constants, given by $\tilde{%
\kappa}_{4}:=(16\pi \tilde{G}_{4})^{-1}=4!\,Vol(K_{3})(c_{1}+c_{2}\Lambda
_{3})$ and $\Lambda _{4}=-3\frac{5c_{0}+c_{1}\Lambda _{3}}{%
c_{1}+c_{2}\Lambda _{3}}$. Here, $Vol(K_{3})$ stands for the volume of $%
K_{3} $. Then, the remaining field equations, corresponding to the
projection along $K_{3}$, generically reduce to an additional scalar
condition on the four-dimensional geometry, fixing the Euler density of $%
M_{4}$ to be a constant, i.e., 
\begin{gather}
(c_{2}+3c_{3}\Lambda _{3})\left( R^{\mu \nu \lambda \rho }R_{\mu \nu \lambda
\rho }-4R^{\mu \nu }R_{\mu \nu }+R^{2}\right) +  \notag \\
24\left[ \frac{c_{0}(5\Lambda _{3}c_{2}-15c_{1})-\Lambda
_{3}c_{1}(5c_{1}-\Lambda _{3}c_{2})}{c_{1}+c_{2}\Lambda _{3}}\right] =0\ .
\label{Euler4}
\end{gather}%
This equation conflicts with most solutions of the Einstein equations. In
particular, for the EGB theory ($c_{3}=0$) the four-dimensional
gravitational field becomes overconstrained, so that even the
Schwarzschild-(A)dS solution is excluded. In the presence of (\ref{Euler4}),
spontaneous compactifications have been found \cite{M-H+DA}. However, the
obstruction on the four-dimensional metric imposed by (\ref{Euler4}) can be
eliminated when the cubic term is switched on, provided the curvature radius
of $K_{3}$ is fixed as $\Lambda _{3}=-c_{2}/(3c_{3})$, for the class of
theories whose couplings are related such that the square bracket of (\ref%
{Euler4}) vanishes. Since the cubic term must be present in order for the
above mechanism to work, the coupling $c_{3}$ can be regarded as the overall
factor of the action, so that it is useful to work with the rescaled
couplings $\tilde{c}_{i}=c_{i}/c_{3}$.

In sum, the class of theories given by 
\begin{equation}
\tilde{c}_{0}=\frac{\tilde{c}_{1}\tilde{c}_{2}(15\tilde{c}_{1}-\tilde{c}%
_{2}^{2})}{15(9\tilde{c}_{1}+\tilde{c}_{2}^{2})}\ ,
\label{class of theories 7D}
\end{equation}%
admits spontaneous compactifications in vacuum where the four-dimensional
geometry fulfills the Einstein field equations in vacuum with neither
corrections nor further constraints, and the effect of the quadratic and
cubic terms just amounts to a redefinition of the Newton and cosmological
constants, which are given by 
\begin{equation}
\tilde{\kappa}_{4}=4!\,Vol(K_{3})c_{3}(\tilde{c}_{1}-\frac{\tilde{c}_{2}^{2}%
}{3})\ ;\ \Lambda _{4}=-\frac{6\tilde{c}_{1}\tilde{c}_{2}}{9\tilde{c}_{1}+%
\tilde{c}_{2}^{2}}\ ,  \label{4D-shifted Newton Constant+Eff Lambda 4}
\end{equation}%
respectively, and the curvature radius of the extra dimensions turns out to
be fixed only in terms of the rescaled Gauss-Bonnet coupling, $\Lambda _{3}=-%
\tilde{c}_{2}/3$.

This clearly differs from the spontaneous compactification of the Einstein
theory which fixes both, $\Lambda _{3}$ and $\Lambda _{4}$ to be
proportional to the cosmological constant in seven dimensions, preventing
the compatibility of a tiny four-dimensional cosmological constant with
small enough extra dimensions. Remarkably, the obstruction appearing in GR
can be surmounted in the present framework, since the induced cosmological
constant in four dimensions (\ref{4D-shifted Newton Constant+Eff Lambda 4})
depends on two parameters. Hence, in this scenario it is possible to
accommodate consistently a tiny cosmological constant in four dimensions
with a sufficiently small compact internal manifold. The condition $|\Lambda
_{4}/\Lambda _{3}|\ll 1$ amounts to requiring $|9\tilde{c}_{1}/\tilde{c}%
_{2}^{2}|\ll 1$, from which it is apparent that the higher curvature terms
cannot be regarded as small corrections of GR in higher dimensions. Note
that it is also possible to have $\Lambda _{3}$ and $\Lambda _{4}$ of
opposite signs without the need of introducing matter fields.

Note that, the effective four-dimensional Newton constant $\tilde{\kappa}%
_{4} $ in Eq. (\ref{4D-shifted Newton Constant+Eff Lambda 4}) acquires a
shift as compared with the standard value $\kappa _{4}$ obtained from
Einstein's theory, which in our conventions reads $\kappa
_{4}=4!\,Vol(K_{3})c_{1}$. Remarkably, the positivity of $\tilde{\kappa}_{4}$
can be guaranteed even if the standard Newton constant in higher dimensions,
given by $c_{1}=(4!16\pi G_{7})^{-1}$, is negative or even vanishing. In
other words, it is possible to recover GR in four dimensions from a
seven-dimensional theory even without Einstein-Hilbert term. This effect
opens up new interesting scenarios where a maximally symmetric solution in
higher dimensions could decay into the spontaneously compactified spacetime
either by tunneling or through a gravitational analogue of ghost
condensation. Indeed, we now show that this is what occurs when both the
extra dimensions and the four-dimensional cosmological constant are required
to be to be small.

\section{Maximally symmetric spacetimes versus spontaneous compactifications.%
}

The presence of cubic terms in the action (\ref{Action7}) allows the
existence of up to three maximally symmetric solutions, and each of them, in
principle, can be regarded as an uncompactified ground state. Indeed, in the
case of constant curvature spacetimes, $R^{AB}=\lambda_{7}e^{A}e^{B}$, the
field equations reduce to a cubic polynomial in $\lambda_{7}$ given by 
\begin{equation}
P(\lambda_{7}):=\lambda_{7}^{3}+\tilde{c}_{2}\lambda_{7}^{2}+\tilde{c}%
_{1}\lambda_{7}+\tilde{c}_{0}=0\ ,  \label{PdB}
\end{equation}
and hence the theory admits at most three possible maximally symmetric
ground states whose radii are determined by the real roots $\bar{\lambda}%
_{7} $ of (\ref{PdB}).

Note that generically, the seven-dimensional Newton constant is not
determined by the coefficient $c_{1}$ in front of the Einstein-Hilbert term,
which could even be absent from the very beginning, since the linearized
field equations around any of the maximally symmetric ground states of
curvature $\bar{\lambda}_{7}$, reduce to the standard Fierz-Pauli equation
but with a different overall factor. This can be seen linearizing the field
equations, expressing the vielbein as $e^{A}=\bar{e}^{A}+\delta e^{A}$,
where $\bar{e}^{A}$\ stands for the chosen background. The linearized field
equations then read 
\begin{equation*}
c_{3}P^{\prime }(\bar{\lambda}_{7})\ \epsilon _{A_{1}...A_{6}B}\ \delta
\left( R^{A_{1}A_{2}}-\!\bar{\lambda}_{7}e^{A_{1}}e^{A_{2}}\right) \bar{e}%
^{A_{3}}...\bar{e}^{A_{6}}\!\!=\!0\ ,
\end{equation*}%
and the corresponding Newton constant is given by 
\begin{equation}
\bar{\kappa}_{7}:=\frac{1}{16\pi \bar{G}_{7}}=4!c_{3}P^{\prime }(\bar{\lambda%
}_{7})\ .  \label{7d Newton}
\end{equation}%
Hence, each ground state possesses its own Newton constant that depends on $%
c_{3}$ and the slope of the polynomial (\ref{PdB}) evaluated at the
corresponding root. Thus, the linearized field equations acquire support
provided the expansion is performed around a nondegenerate root. Since the
sign of the second variation of the effective action that defines the
graviton propagator is determined by $\bar{\kappa}_{7}$, for different
uncompactified ground states, the graviton could behave as a particle or as
a ghost, as occurs for the EGB theory \cite{BD}.

It is worth remarking that the maximally symmetric spacetimes for which the
graviton behaves as a ghost should not be discarded from scratch. Indeed,
from the point of view of dimensional reduction, one might expect that they
correspond to a sort of false vacua which could decay to the spontaneously
compactified spacetime, where the graviton has a well-defined propagator,
through a gravitational analogue of ghost condensation (see Ref. \cite{Ghost}%
). In this case, the transition, instead of being driven by an additional
scalar field, is triggered by the higher dimensional graviton in vacuum.
Note that even those vacua whose seven-dimensional graviton does behave as a
particle could still be metastable against decay into the compactified
spacetime by tunneling.

Both possibilities open up new interesting scenarios which do not occur for
spontaneous compactifications of (super) gravity based on the
Einstein-Hilbert action in higher dimensions, which nevertheless could be
compatible with some more recent alternatives to compactification \cite%
{Alternatives}. Note that for a theory that possesses a maximally
(super)symmetric vacuum with well-defined propagators around it, it is not
simple to explain why the theory does not prefer this configuration as the
ground state, instead of choosing a compactified spacetime. In the present
framework, considering a theory that does not possesses a suitable
uncompactified ground state is welcome, since it could naturally provide a
dynamical mechanism of spontaneous dimensional reduction (see also \cite{HTZ}%
).

Let us discuss the possible scenarios within the class of theories defined
by (\ref{class of theories 7D}), which admits spontaneous compactifications
consistent with GR in vacuum. The arbitrariness in the choice of the
coupling constants for this class of theories is further restricted
requiring a few conditions compatible with a realistic four-dimensional
picture. Thus, requiring both a tiny (preferably positive) four-dimensional
cosmological constant and small enough extra dimensions implies $\left\vert
\Lambda _{4}\right\vert \ll \left\vert \Lambda _{3}\right\vert $, which in
terms of the coupling constants reads $9\left\vert \tilde{c}_{1}\right\vert
\ll \tilde{c}_{2}^{2}$. Consequently, since the effective Newton constant in
four dimensions $\tilde{\kappa}_{4}$ has to be positive, Eq. (\ref%
{4D-shifted Newton Constant+Eff Lambda 4}) implies that the overall factor
in front of the action must be negative. In other words, the condition $%
c_{3}<0$ is necessary in order to have a well-defined propagator for the
graviton in four dimensions.

Albeit it is not necessary, hereafter we restrict to the case $\Lambda
_{3}>0 $ (i.e. $\tilde{c}_{2}<0$) in order to allow the presence of
non-Abelian gauge fields. Note that $c_{2}>0$ is compatible with string
theory \cite{BD}. The analysis then splits into three cases according to the
sign of $\tilde{c}_{1}$, which due to the above requirements, coincides with
the sign of $\Lambda _{4}$:

$\Lambda _{4}<0$: The polynomial (\ref{PdB}) has positive slope at its
unique real root which is simple and positive. Thus, there is a single
maximally symmetric solution of positive curvature ($dS_{7}$), and since $%
\bar{\kappa}_{7}<0$\ (see Eq. (\ref{7d Newton})), the seven-dimensional
graviton behaves like a ghost. The spontaneous compactification has a
well-defined graviton in four dimensions (see Eq. (\ref{4D-shifted Newton
Constant+Eff Lambda 4})). This suggests that $dS_{7}$ could spontaneously
decay into the less symmetric solution, $AdS_{4}\times S^{3}$, through the
gravitational analogue of ghost condensation.

$\Lambda_{4}=0$: The polynomial $P(\lambda_{7})$ has two real roots, and it
has positive slope at the simple one which is positive. Thus, one of the
maximally symmetric vacua corresponds to $dS_{7}$, which is expected to
decay through gravitational ghost condensation into the compactified vacuum, 
$\mathbb{M}_{4}\times S^{3}$, having a well-defined propagator for the
graviton. The doubly degenerate root vanishes (i.e., $\bar{\lambda}_{7}=0$
and $\bar{\kappa}_{7}=0$), so that the maximally symmetric solution is
Minkowski spacetime around which the linearized equation for graviton has no
support. Thus, since the propagator is ill-defined, flat spacetime is
expected to be a sort of false vacuum. This scenario is similar to the one
in Ref. \cite{HTZ}, where it is shown that in order to have propagation for
the graviton, the spatial components of the curvature cannot be small.
Hence, propagating deviations around flat spacetime are nonlocal and would
require too much energy, since they are nonperturbative. In this case the
spontaneous compactified spacetime, $\mathbb{M}_{4}\times S^{3}$, is
expected to be preferred since it admits a well-defined low energy limit.

$\Lambda _{4}>0$: $P(\lambda _{7})$ has three real roots, $\lambda
_{7}^{(1)}<0<\lambda _{7}^{(2)}<\lambda _{7}^{(3)}$. As the slope of the
polynomial at $\lambda _{7}^{(1)}$ and $\lambda _{7}^{(3)}$ is positive, the
seven-dimensional graviton behaves as a ghost around the corresponding
maximally symmetric $AdS_{7}$ and $dS_{7}$ spacetimes, respectively, which
would decay into the spontaneous compactification $dS_{4}\times S^{3}$.
Since the slope of the polynomial at the remaining root is negative, the
corresponding Newton constant becomes positive, and so the graviton in seven
dimensions acquires a well-defined propagator around $dS_{7}$ with curvature 
$\lambda _{7}^{(2)}$. Nonetheless, this seven-dimensional vacuum appears to
be metastable against decay through tunneling into the compactified
configuration $dS_{4}\times S^{3}$. A crude estimate of the tunneling rate
can be obtained in the semiclassical approximation through the difference of
the Euclidean version of the action (\ref{Action7}) evaluated on the
corresponding Eucildean continuations, $S^{7}$ and $S^{4}\times S^{3}$,
respectively.

In the case of $S^{4}\times S^{3}$, for theories with $x:=9\tilde{c}_{1}%
\tilde{c}_{2}^{-2}=(2\Lambda_{3}/\Lambda_{4}-1)^{-1}\ll1$ (i.e., $\Lambda
_{4}\ll\Lambda_{3}$), the Euclidean action, up to an overall positive
factor, becomes $I_{4,3}\simeq-c_{3}x^{-1}$; while for $S^{7}$, of curvature
given by $\lambda_{7}^{(2)}=(\tilde{c}_{1}/15)^{1/2}+\mathcal{O}\left(
x\right) $, the action is given by $I_{7}\simeq-c_{3}x^{-3/4}$. Since in our
conventions the most likely vacuum is the one with the larger Euclidean
action, for a realistic picture (i.e. for small $x$), the seven-dimensional
de Sitter solution turns out to be metastable against tunneling into the
spontaneously compactified vacuum $dS_{4}\times S^{3}$, because the latter
clearly has a larger Euclidean action.

Note that this is not a possibility for the Einstein theory with positive $%
\Lambda_{7}$ in vacuum, since not only $\Lambda_{3}$\ and $\Lambda_{4}$ are
of the same order as $\Lambda_{7}$, but the tunneling mechanism goes in the
opposite way, since $dS_{7}$ turns out to be more likely than $dS_{4}\times
S^{3}$. Indeed, as the on-shell value of the Euclidean action is given by $%
I_{EH}(M_{7})=(20\pi G_{7})^{-1}\Lambda_{7}Vol(M_{7})$, and as the seven
sphere has a better distributed volume than $S^{4}\times S^{3}$, one obtains
that $I_{EH}(S^{7})/I_{EH}(S^{4}\times S^{3})=3^{3/2}/4>1$. Therefore, GR in
vacuum prefers the uncompactified spacetime. In our framework as the action
possesses additional fundamental constants, the tunneling mechanism acquires
a control parameter $x$, which for a small enough value, allows a concrete
realization of spontaneous breakdown of the vacuum symmetry from $SO(7,1)$
to $SO(4,1)\times SO(4)$.

In sum, for the class of theories admitting spontaneous compactifications
consistently with GR in vacuum, defined by (\ref{class of theories 7D}),
once required to be compatible with a reasonable four-dimensional picture,
the spontaneously compactified vacuum turns out to be preferred with respect
to the maximally symmetric spacetime, since the latter would decay either by
the gravitational analogue of ghost condensation or by tunneling.

\section{Gauge fields with shifted coupling.}

Since the theory (\ref{Action7}) is generally covariant the massless gauge
fields are guaranteed to be gauge invariant. Switching on only the massless
modes for the gauge field, the field equations reduce to Yang-Mills at the
linearized level, i.e.%
\begin{equation}
4(3c_{1}+c_{2}\Lambda _{3})\bar{\nabla}^{\mu }\partial _{\lbrack \mu }A_{\nu
]}^{(a)}=0\ ,  \label{YM}
\end{equation}%
Hence, as $\Lambda _{3}=-c_{2}/(3c_{3})$, the gauge coupling is shifted by $%
c_{1}\rightarrow c_{1}-c_{2}^{2}/(9c_{3})$, compared with Weinberg's formula
obtained from compactification of GR \cite{Weinberg}. Note that the gauge
coupling is neither proportional to the standard nor the effective Newton
constant (\ref{4D-shifted Newton Constant+Eff Lambda 4}). The gauge field
possesses a well-defined propagator provided the effective gauge coupling is
positive, and it is reassuring to verify that this is the case for the
realistic scenarios discussed above with $|\Lambda _{4}/\Lambda _{3}|\ll 1$
and $\tilde{\kappa}_{4}>0$.

As occurs for the compactifications of GR, the analysis of the scalar field
propagator requires special attention. This, as well as how the Kaluza-Klein
tower differs from the one obtained from standard (super)gravity in higher
dimensions, is left for future research. It is also worth mentioning that
the fact that Lovelock theories admit propagating degrees of freedom for
torsion in vacuum \cite{TZ} may help to obtain realistic chiral fermions
from spontaneous compactifications \cite{Witten}.

The mechanism that leads to spontaneous compactifications in vacuum
consistent with GR in four dimensions actually carries through in $D\geq 7$
dimensions. It would be interesting to explore compactifications on product
spaces, since as it occurs in standard supergravity, this may improve the
stability of the solution, naturally suggesting time-dependent
compactifications \cite{Oh}. Previous results \cite{Naty} indicate that
within our framework the compact manifold may shrink in time (related
cosmological models have been studied in \cite{Naty2}). For theories with
quartic or higher powers of the curvature it is possible to have more than
one possible compactification radius, enlarging the class of theories that
admits suitable spontaneous compactifications. Note that, for theories with
quartic or higher even powers of the curvature, a maximally symmetric vacuum
may not even exist, suggesting that the ground state must be compactified.
This is discussed in a forthcoming publication.

\acknowledgements We thank G. Giribet, G. Kunstatter, D. Marolf and J. Oliva
for helpful comments. This work was partially funded by FONDECYT grants N%
${{}^o}$
1051056, 1061291, 1071125, 1085322, 3070057, 11080056, 1095098 and UACH-DID
grant N%
${{}^o}$%
. S-2009-57. Centro de Estudios Cient\'{\i}ficos (CECS) is funded by the
Chilean Government through the Millennium Science Initiative and the Centers
of Excellence Base Financing Program of CONICYT. CECS is also supported by a
group of private companies which at present includes Antofagasta Minerals,
Arauco, Empresas CMPC, Indura, Naviera Ultragas and Telef\'{o}nica del Sur.
CIN is funded by CONICYT and Gobierno Regional de Los R\'{\i}os.

\end{document}